\documentclass[fleqn,usenatbib]{mnras}
\newcommand{\RN}[1]{\uppercase\expandafter{\romannumeral #1\relax}}

\usepackage{newtxtext,newtxmath}
\usepackage{lineno}
\usepackage{soul} % \st{}删除线
\usepackage{color} % \textcolor{red}{}
\usepackage[T1]{fontenc}

\DeclareRobustCommand{\VAN}[3]{#2}
\let\VANthebibliography\thebibliography
\def\thebibliography{\DeclareRobustCommand{\VAN}[3]{##3}\VANthebibliography}

\usepackage[figuresright]{rotating}
\usepackage{caption}
\usepackage{graphicx}	% Including figure files
\usepackage{amsmath}	% Advanced maths commands
\title[]{Discovery of a new IW And-type dwarf nova with both tilted disk and tidal instability}

\author[Sun et al.]{
Yongkang Sun,$^{1,2}$
Xin Li,$^{3}$
Qige Ao,$^{4,2}$
Wenyuan Cui,$^{5}$\thanks{E-mail: wenyuancui@126.com, cuiwenyuan@hebtu.edu.cn}
Bowen Zhang,$^{1,2}$
Yang Huang,$^{2,1}$
Jianrong Shi,$^{1,2}$
\newauthor
Linlin Li$^{5}$, and Jifeng Liu$^{1,2}$
\\
$^{1}$Key Laboratory of Optical Astronomy, National Astronomical Observatories, Chinese Academy of Sciences, Beijing 100012, People's Republic of China\\
$^{2}$School of Astronomy and Space Science, University of Chinese Academy of Sciences, Beijing 100049, People's Republic of China\\
$^{3}$Beijing Planetarium, Beijing Academy of Sciences and Technology, Beijing 100044, People's Republic of China\\
$^{4}$Shanghai Astronomical Observatory, Chinese Academy of Sciences, 80 Nandan Road, Shanghai 200030, People's Republic of China\\
$^{5}$College of Physics, Hebei Normal University, Shijiazhuang 050024, People's Republic of China\\
}

\date{Accepted XXX. Received YYY; in original form ZZZ}

\pubyear{2023}

\begin{document}
\label{firstpage}
\pagerange{\pageref{firstpage}--\pageref{lastpage}}
\maketitle

\begin{abstract}
IW And-type dwarf novae are anomalous Z Cam stars featured with outbursts happening during standstill states, which are not expected in the standard disk instability model. The physical mechanisms for these variations remain unclear. In this study, we report the discovery of a new candidate IW And-type dwarf nova J0652+2436, identified with its frequent outbursts from the slowly rising standstill states. Luckily, the TESS observations during a long standstill state and the earlier K2 observations give a chance to find the orbital and negative superhump period in the light curve of J0652+2436, allowing the measurement of its mass ratio of 0.366. This mass ratio is marginally possible for the tidal instability to set in according to previous SPH simulations. Thus, we propose that the outbursts in J0652+2436 are likely to be caused by the growing accretion disk during standstills, in favor of the previous hypothesis of the mechanisms lying in all IW And stars. We conclude that J0652+2436 might be the first IW And star with both a precessing tilted disk and tidal instability, which will be an important laboratory for studying the accretion disk dynamics and help understand IW And phenomenon.

\end{abstract}


%%%%%%%%%%%%%%%%%%%%%%%%%%%%%%%%%%%%%%%%%%%%%%%%%%

%%%%%%%%%%%%%%%%% BODY OF PAPER %%%%%%%%%%%%%%%%%%

\section{Introduction}

Dwarf novae (DNe), one type of cataclysmic variables (CVs) , show regular outbursts with amplitude typically between 2-5 mag \citep{warner_cataclysmic_1995}, which are explained by the disk instability model (DIM) \citep[e.g.][]{hameury2020review}. The three main subtypes of DNe, U Gem-, SU UMa-, and Z Cam-type, are defined by their light curve characteristics. U Gem stars only show normal outbursts. SU UMa stars are short-period systems that show larger and longer outbursts in addition to normal outbursts, called superoutbursts. The superoutbursts are caused by the thermal-tidal instability when the disk radius reaches the 3:1 resonance radius and the disk becomes eccentric and precessing, during which positive superhumps are produced in the light curve \citep[e.g.][]{osaki_dwarf-nova_1996}. The period of positive superhumps is longer than the orbital period by a few percent, from which the mass ratio can be derived \citep{kato2009survey}. Z Cam stars occasionally enter standstill states after an outburst maximum, during which the brightness is fainter by typically 1 mag than the maximum. These stars usually have relatively short outburst cycles from 10 to 30\,d and do not spend much time at the minimum brightness \citep{ZCamstarsinthe21}. The accretion rate of Z Cam stars is near the upper critical point of the disk-instability zone \citep{dubus2018testing}, so the disk occasionally switches from an unstable DN-like outburst state to a stable standstill state similar to nova-like (NL) stars. The duration of the standstills of different stars is largely variable. 

For normal Z Cam stars, the standstills always start at the decline phase of an outburst, and end with a drop in brightness \citep{hameury2014anomalous}, while a subset of Z Cam stars are found to have standstills terminated by an outburst instead of a decline, and often show consecutive `heart-beat'-like outbursts (each outburst followed by a dip). These stars are called anomalous Z Cam stars, or IW And-type stars \citep{kato2019three}. However, outbursts occurring at the end of standstills are not expected in the traditional DIM model because the disk in high accretion rate is fully ionized and already in a stable state \citep{hameury2020review}. The origin of the outbursts happening during high states remains elusive. 

By now, other explanations for the IW And-type phenomenon have been put forward, including models involving mass-transfer outbursts from the secondary star \citep{hameury2014anomalous} or involving a tilted disk \citep{kimura2020thermal}. Unfortunately, in particular sources like HO Pup, no clear evidence of mass transfer variations were found \citep{lee2021ho}. The mechanism generating mass-transfer outbursts are also elusive. The latter model, which assumes the accretion stream flows into the inner part of a tilted disk, is also challenged. 

When the disk is tilted, negative superhumps can be detected in the light curve with a period of a few percent shorter than the orbital period. The negative superhump is thought to be a beat frequency of a shifting hotspot around the plane of a tilted, retrogradely precessing disk and the orbital frequency. The difference between the negative superhump period and orbital period is related to the mass ratio \citep{thomas_emergence_2015}. 

In the tilted-disk model, a wide range of different types of outbursts can be produced, some of them reproduce similar light curves to those seen in several known IW And-type stars.  
 \citet{kimura2020kic} discovered that the IW And star KIC 9406652 possesses negative superhumps, but the change of the superhump frequency indicates the expansion of the disk, contradictory to the latter model which expects the outer and cooler part of the disk will contract \citep{kimura2021kic}.  In addition, \citet{kato2022analysis} found no negative superhumps in other IW And-type stars, also making the tilted-disk model less effective. 

\citet{kato2019three} found that the SU UMa-type star NY Ser shows superoutbursts during standstills that can be explained by the disk radius exceeding the 3:1 resonance radius. They propose that the disks in IW And stars may also experience the same process in which the disks gradually expand and finally trigger the tidal instability, and the instability spreads from the outer to the inner part of the disks, generating an outburst. 

LAMOST J065237.19+243622.1 (henceforth J0652+2436) has been identified as a CV in the spectral data mining from Large Sky Area Multi-Object Fiber Spectroscopic Telescope (LAMOST) \citep{zhao2012lamost,cui2012large} Data Release 6 (DR6) database \citep{sun2021catalog}. In this work, we analyze the photometry and spectral data of J0652+2436 and classify it as a new IW And-type star. We collect the photometry data from survey programs, ZTF, ASAS-SN, K2 and TESS,  study the profile of light curve and identify the orbital period and superhump period. By comparing the properties of this object with those of other known IW And stars, the peculiarity of J0652+2436 and the high occurrence rate of the IW And phenomenon among Z Cam population are revealed.

The structure of this paper is as follows. In Section 2, we describe the observational data. The analysis and results are given in Section 3. We discuss the properties of J0652+2436 and other related stars in Section 4. Finally, we summarize in Section 5.

\begin{figure*}
	\includegraphics[width=2.0\columnwidth]{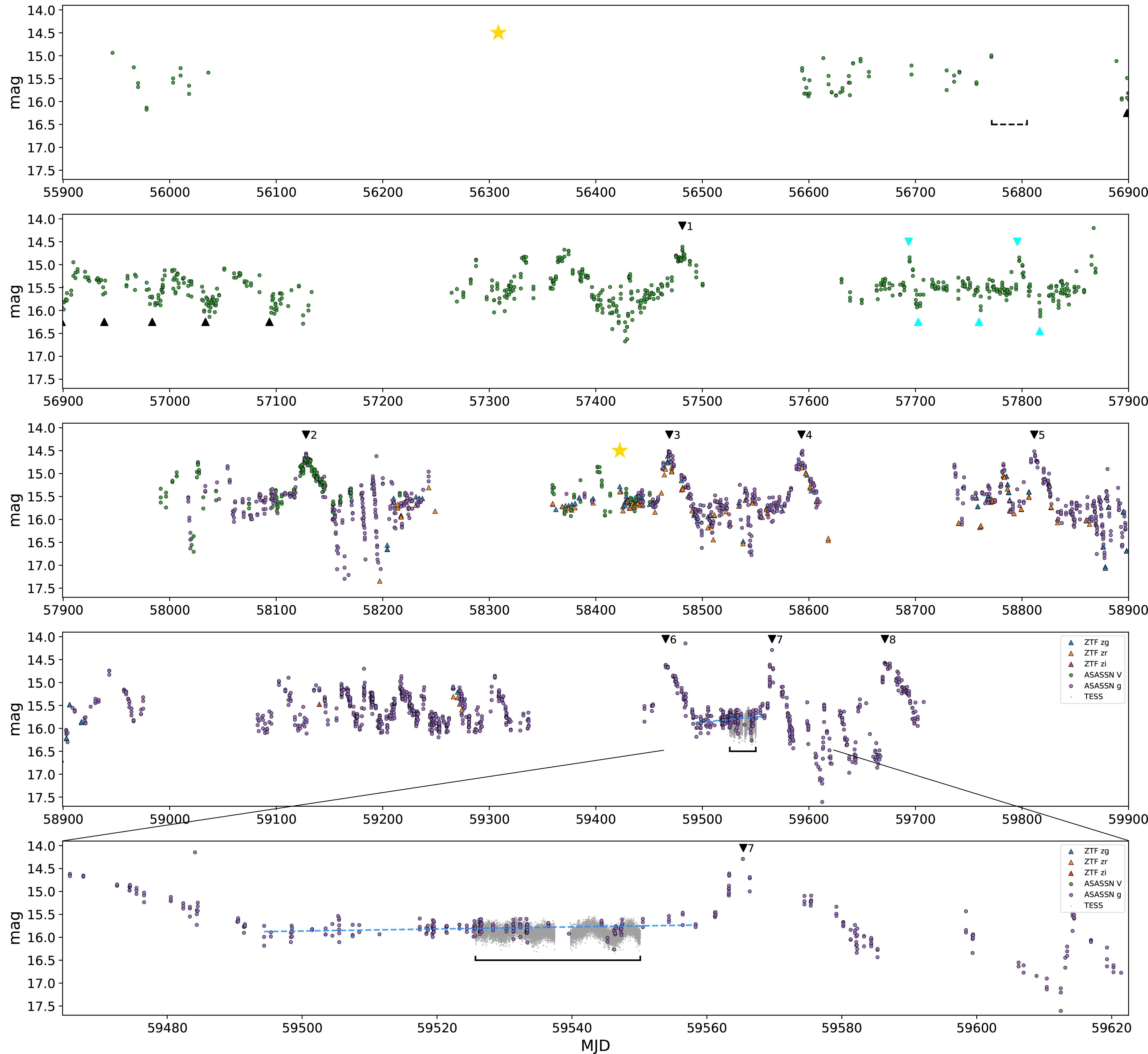}
    \caption{Photometric data of J0652+2436 from ZTF, ASAS-SN, and TESS observations are shown here. Panels 1-4 are continuous and Pannel 5 is a zoom-in of the vicinity of the TESS observation. The MJD values are calculated from HJD - 2400000.5. Yellow stars indicate the time when two LAMOST spectra are taken (see Section \ref{sec:LAMOST}). The coverage of K2 data is indicated by the black dashed horizontal line between MJD 56772 to 56805. The coverage of TESS data is indicated by the black solid horizontal line in the fourth and fifth panels. The flux in the TESS data is converted to magnitude and adjusted to the level matching the ASAS-SN data so that the long-timescale variation of the TESS and ASAS-SN data can be compared. In the last two panels}, the blue dashed line indicates the liner fit of the slowly increased brightness.
    \label{fig:lc1}
\end{figure*}

\section{Data and Observations}
\subsection{Photometry data}

We obtained the photometric data of J0652+2436 from Zwicky Transient Facility (ZTF) DR12 \citep{PASP..131a8002B} catalog\footnote{https://irsa.ipac.caltech.edu/Missions/ztf.html} and All-Sky Automated Survey for Supernovae (ASAS-SN) \citep{shappee2014man, kochanek2017all} sky patrol. The ASAS-SN data was required from the program website\footnote{https://asas-sn.osu.edu/} in the aperture photometry mode. The short-term photometric data of Kepler K2 \citep*{howell2014k2} and Transiting Exoplanet Survey Satellite (TESS) \citep{2014SPIE.9143E..20R} observations are obtained through the \textsc{lightkurve} package \citep{2018ascl.soft12013L} in Python. 

J0652+2436 was observed from April to May 2014 during K2 Mission with the id ktwo202061808. We use the 1800s-cadence K2 self-flat-fielding (SFF) light curve, which was extracted by \citet{2014PASP..126..948V} with corrections for the systematics caused by the failure of reaction wheels of Kepler satellite. By dividing the data into segments and calculating the Lomb-Scagle periodogram \citep{lomb1976least,scargle1982studies} for each of them, we find that the maximum peaks of the periodograms show a more stable peak period in the K2SFF light curve than the original K2 light curve. 

J0652+2436 was monitored by TESS for more than 24\,d in TESS sector 45 in 2021, with a 2.3\,d gap in the middle which divides the light curve into two parts. We use the 2-minute cadence light curve processed by the Science Processing Operations Center (SPOC) pipeline \citep{jenkins2016tess}. In searching periodic signals in TESS light curve, we use the presearch data conditioning simple aperture photometry (PDCSAP) flux in our analysis.

\subsection{LAMOST spectra}\label{sec:LAMOST}

We downloaded low-resolution spectra (R \textasciitilde\ 1800 at 5500\,\AA) of J0652+2436 from the LAMOST DR9 online database \footnote{http://www.lamost.org/dr9/v1.0/}. The wavelength coverage is from 3700\,\AA\ to 9000\,\AA. There are two low-resolution spectra which are taken on MJDs 56309 and 58423, respectively. The exposure time of each spectrum is 1800\,s and 4500\,s, respectively.

\subsection{Xinglong 2.16m spectroscopy}\label{sec:rv}
We obtained time-series spectra using the BFOSC spectrograph equipped on Xinglong 2.16m telescope \citep{fan2016xinglong} on 2022 November 15. The G8 grism and 1.8" slit were selected based on the seeing condition. We acquired six spectra with an exposure time of 1320 s for each, R \textasciitilde\ 480 and wavelength range of 6140 - 8270\,\AA. Bias and flat corrections as well as wavelength calibration were performed using IRAF \citep{tody1986iraf,tody1993iraf}.
%(see Figure~\ref{fig:216})
% MJD59898

\section{Analysis}

\subsection{The outbursts}\label{sec:outbursts}

Figure~\ref{fig:lc1} shows the ZTF g/r/i-bands, ASAS-SN V/g-band, and TESS photometry data of J0652+2436. For ASAS-SN data, we remove the data points of which the magnitudes are higher than the limit magnitude or have a value of 99.990. The ASAS-SN observations have a much wider temporal coverage. As seen in the long-term light curve, the star has considerable variability. It mainly shows outbursts with amplitudes of \textasciitilde1 mag occurring during an intermediate-bright standstill state at \textasciitilde16 mag level, while there are apparently different outburst patterns.

Between MJD 56899 to 57100, the star shows continuous small outbursts. We denote the time of the minima of the outbursts by upward black triangles below the data-points after we take out this segment alone and bin them by 5\,d. The intervals between the successive minima are 40, 45, 50, and 60\,d, respectively. The minima is at an level of about 15.9 mag in V band, and the maxima reach 15.1 mag. 

From MJD 57690 to 57850, the star shows two small outbursts and three dips from a stable high state with V-band magnitudes of around 15.5 mag, as denoted by cyan triangles. The interval between the two outbursts is about 100\,d, and the intervals between the three dips are both \textasciitilde60\,d. The maximum brightness of the outbursts in V band is about 14.8 mag and the outburst duration is about 10\,d. The minima of dips reaches 16.1 mag. The standstill states between the dips are stable with a level of about 15.5 mag. These outbursts followed by dips resemble the 'heart-beat'-shape outbursts seen in known IW And-type stars. Although it is a little unexpected that no outburst was recorded before the second dip, the \textasciitilde60\,d intervals are similar to those reported in IW And (recurrence time of \textasciitilde40 d), V513 Cas (recurrence time of 40-50 d) and IM Eri (recurrence time can reach above 60 d) \citep{kato2019three}.

From MJD 59100 to 59350, the star shows another phase of continuous small outbursts occurring during the intermediate-bright state. The recurrence time of this phase is \textasciitilde28 d, shorter than that of the outbursts between MJD 56899 to 57100. The minima of this phase is around 15.9 mag in g band, and the maxima is around 15.0 mag, so the outburst amplitudes are similar to those between MJD 56899 and 57100.

Besides, the star shows large outbursts followed by a deep dip to a low state and successive short outbursts arising from the low level with small recurrence time and increasing amplitudes. The similar outburst pattern has been seen in a SU UMa star NY Ser which is discussed in Section~\ref{sec:compare}. The largest eight outbursts are marked by black downward triangles above the data points in the figure and are denoted as outburst 1-8. The star reaches its maximum brightness in these outbursts, which is about 14.5 mag in g band. The lowest brightness is reached at the dips after these outbursts, which is 17.3 mag in g band, despite only one data point recorded being below 17.5 mag. The recurrence time of the successive outbursts arising from the low level is variable.

\begin{figure}
	\includegraphics[width=\columnwidth]{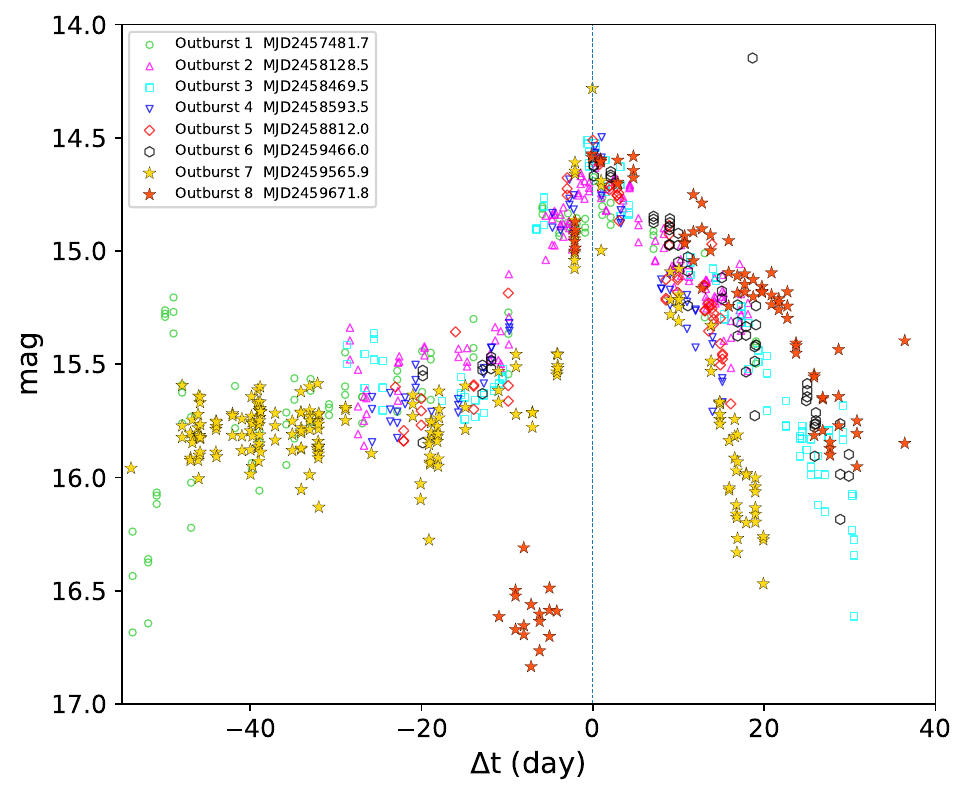}
	\caption{The eight largest outbursts in the long-term light curve from ASAS-SN V (for outbursts 1 and 2) and g (for outbursts 2-8) band data are shown. Different markers are used to indicate each outburst. $\Delta t$ is defined as the time difference from the outburst maximum. The peak MJDs of each outburst are manually shifted to $\Delta t = 0$ and are indicated in the legend.}
	\label{fig:outbursts}
\end{figure}

To compare the profiles of the largest outbursts, we shift the times of each outburst to make their maximum at the same time, defined as $\Delta t=0$. The results are shown in Figure~\ref{fig:outbursts}. Each outburst presents nearly the same peak magnitude, which is 14.5-14.6 mag, despite an outlier in outburst 7. Outbursts 3-5 have a peak value of 14.5 mag, while the other five have slightly lower peak values of about 14.6 mag. All but the eighth outburst are initiated at the higher level of about 15.6 mag. The eighth outburst is, however, initiated at the previous low level of 16.8 mag. For outbursts 1-6, the rise rates are nearly the same, which are around 0.07-0.09 mag per day, and the rises lasted about 14\,d. Different from the previous six outbursts, the rise rates of outbursts 7 and 8 are nearly identical and particularly rapid (0.32 and 0.40 mag per day lasting over 6 d). The decline of the eight outbursts shows different rates, but can mainly be divided into two stages, first a slower and then a faster descent, and the magnitude of the onset of the faster descent phase is nearly the same. 
In summary, large outbursts seen in J0652+2436 are triggered in either the low or high state, and the onset at the high state has the same brightness. The outburst rise stages show two kinds of rates, while the decline shows two stages with the same transition magnitude. The consistency of the maximum brightness of the eight outbursts may indicate that the accretion disk reaches its maximum radius at these peaks.

The presence of low states with minimum light below mag = 17 apart from the standstill states means that J0652+2436 must have a hot and bright accretion disk most of the time. From MJD 59494, the star entered a stable standstill state for about 67\,d, which fits the important characteristic of Z Cam-type dwarf nova. However, this standstill state was then interrupted by an outburst at MJD 59561, which is not expected for a normal Z Cam star because a hot and fully ionized disk should be stable and not produce outbursts under the interpretation of DIM \citep{hameury2014anomalous}. This critical feature is typical for an IW And-type star. 
In addition, the standstill shows an average brightening of 0.0022 mag per day, from \textasciitilde15.88 mag at the start to around \textasciitilde15.73 mag at the end. Large-scale periodic fluctuations can be seen during at this stage, which can be directly seen in the 25-d TESS observation (analyzed in Section~\ref{sec:TESS_lc}).

To summarize, J0652+2436 showed complex outburst patterns. The small outbursts with dips appeared during MJD 57631-57853 and a standstill state terminated by a large outburst makes J0652+2436 similar to known IW And-type dwarf novae. Although diverse light variations have been seen within a single IW And star \citep{kimura2020thermal}, the outbursts that showing different amplitudes, recurrence time and duration, as well as the varying standstills make J0652+2436 somewhat peculiar.

\subsection{Spectral features}
\begin{figure}
	\includegraphics[width=\columnwidth]{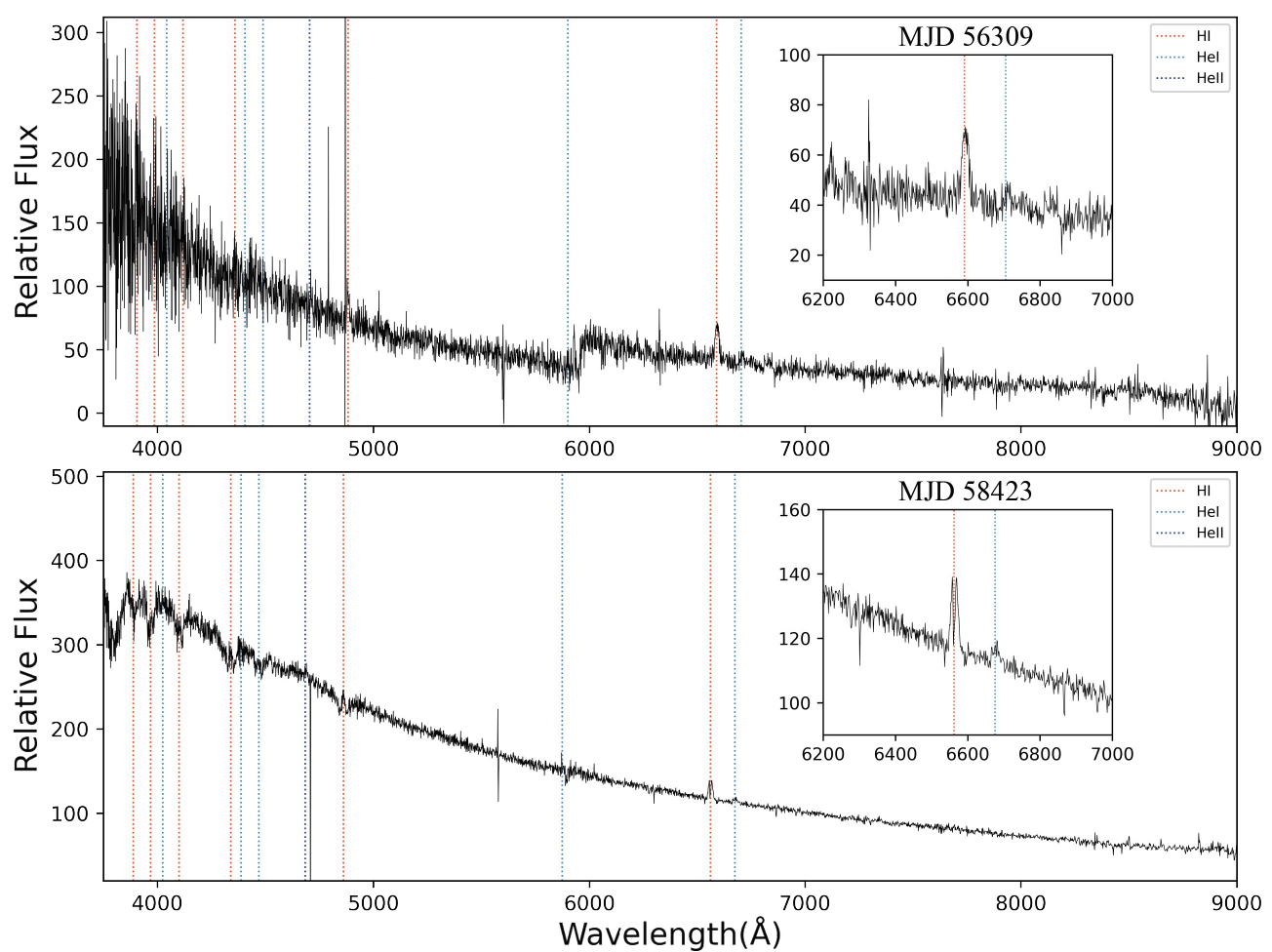}
	\caption{The two LAMOST spectra of J0652+2436 are plotted in the upper and lower panel. The Balmer, HeI, and HeII lines are marked by red, blue, and dark blue lines, respectively. The right side in each panel is the zoom-in view of the H$\alpha$ and He\,{\small\RN{1}} $\lambda$6678 region.}
	\label{fig:LAMOST}
\end{figure}

In Figure~\ref{fig:LAMOST}, we show the two LAMOST spectra of J0652+2436. They both show blue continuums, but no visible low-temperature star feature. So the spectra are dominated by the emission of the accretion disk. The most prominent feature is the broad H$\alpha$ emission, which are present in both spectra. The He\,{\small\RN{1}} $\lambda$6678 emission line also appears, which is typical for CVs. Due to the low signal-to-noise ratio (SNR) of the spectrum at MJD 56309, the H$\beta$ and He\,{\small\RN{2}} $\lambda$4686 emission lines are barely visible in the blue part of the spectrum. The narrow line at the left side of H$\beta$ is probably due to cosmic ray hit. The jump near 6000\,\AA\ is because the red and blue ends of the spectrum are not well spliced. 
The features of blue continuum, Balmer emission cores flanking absorption troughs and He\,{\small\RN{2}} $\lambda$4686 emission line are similar to those of IW And in outburst and standstill state as shown in  \citet{Szkody2013}. The features also consistent with those in the spectrum of IW And star HO Pup taking at an outburst event \citep{lee2021ho} showing Balmer emission cores and He\,{\small\RN{1}} + He {\small\RN{2}} emission. For the spectrum taken at MJD 58423, it can be directly confirmed that was taken during a small outburst. As for the spectrum taken at MJD 56309, it was also likely to be taken during standstill or outburst state as J0652+2436 was in a state producing continuous small-amplitude outburst as that period.

In CVs, the white dwarf can hardly be eclipsed when the inclination is less than 70$^{\circ}$. The fact that there are no eclipses in the light curve and the widths of the H$\alpha$ emission line compared with those of other LAMOST CVs suggest that J0652+2436 is likely to have an intermediate inclination. 

\subsection{Periodical signals}

\begin{figure}
	\includegraphics[width=\columnwidth]{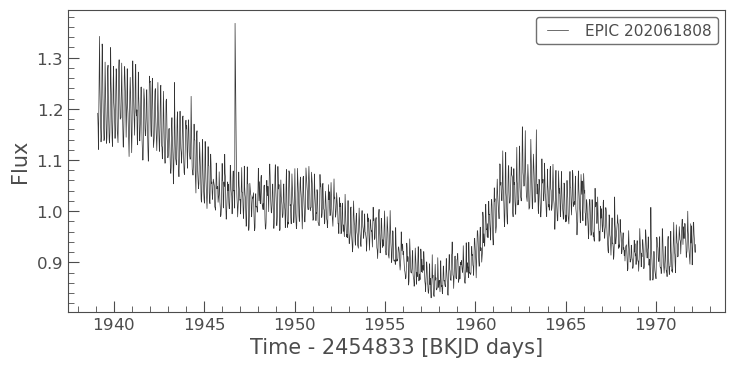}
	\caption{This is the SFF corrected K2 light curve of J0652+2436. The outliers above 5 sigma have been removed from the downloaded data.}
	\label{fig:lck2}
\end{figure}

We perform the periodogram analysis using the Time Series module of the \textsc{astropy} package \citep{astropy:2013,astropy:2018,astropy:2022}.

\subsubsection{K2 light curve}
\begin{figure}
	\includegraphics[width=\columnwidth]{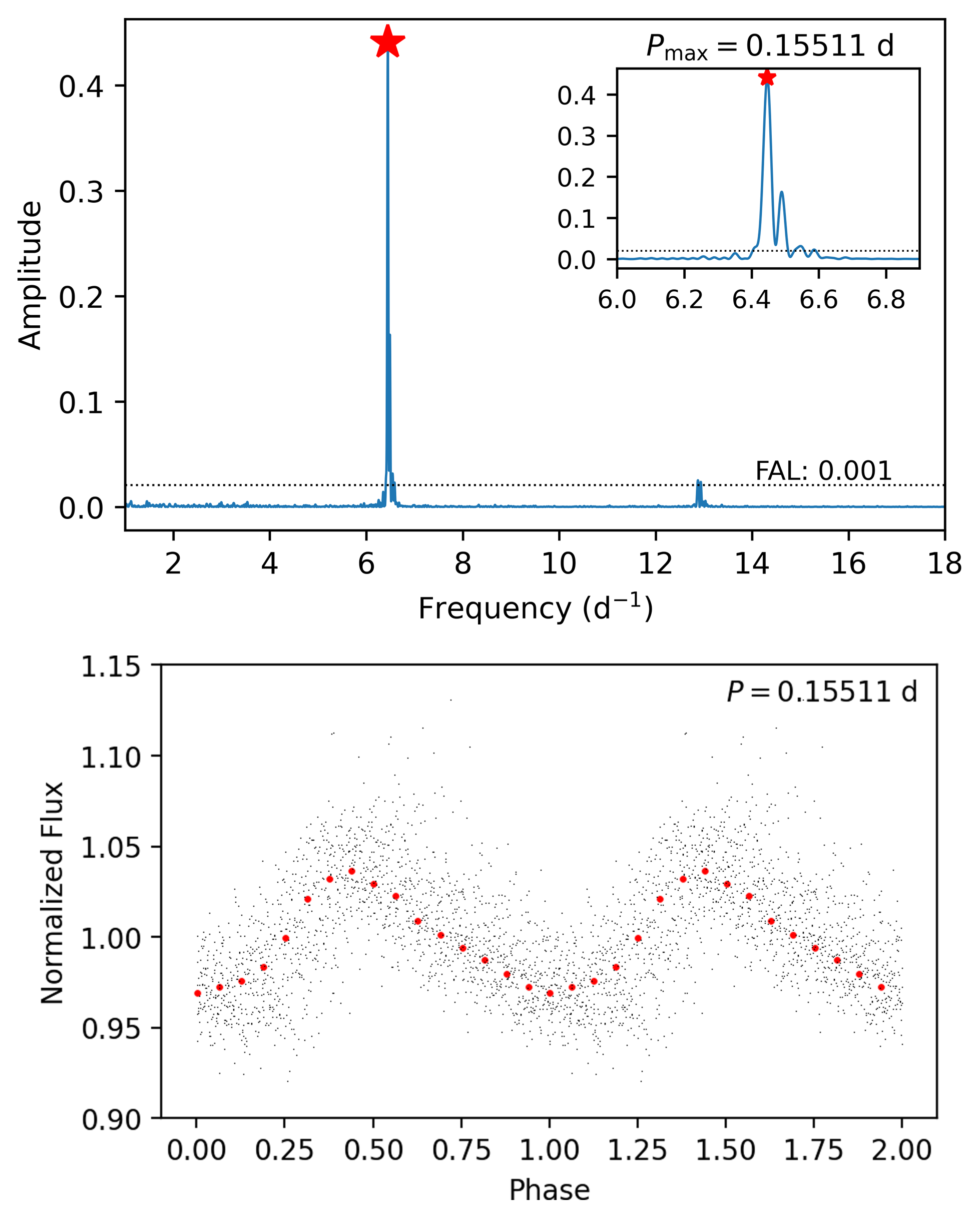}
	\caption{Upper panel: The Lomb-Scargle periodogram of the flattened K2 SFF light curve is calculated. The red star marks the maximum peak which has a period of 0.15511\,d. The black dotted line is 0.001 FAL. The right upper panel is a zoomed view near the maximum peak.
	Lower panel: Phase-folded light curve of the flattened Kepler data is plotted using the period of 0.15511\,d. The gray dots are original data and the red dots are 14-minute binned data. For clarity, two phases are drawn repeatedly.}
	\label{fig:K2_A}
\end{figure}

In order to search periodical signals in the K2SFF light curve (shown in Figure~\ref{fig:lck2}), we first flatten the light curve to remove the large-scale flux change and remove outliers with fluxes above 5 sigmas, after which 1580 valid data points are obtained. The Lomb-Scargle periodogram is then calculated using the \textsc{astropy.timeseries} subpackage, which is shown in the upper panel of Figure~\ref{fig:K2_A}. The frequency corresponding to the maximum peak is 6.4470 $\rm d^{-1}$, or a period of 0.15511\,d. Folded on this period for the flattened light curve, the resulting phase profile (shown in the lower panel of Figure~\ref{fig:K2_A}) deviates from a sinusoidal form, which is also indicated by the existence of higher harmonics in the periodogram. The maximum is located at around phase 0.4, with an amplitude of about 0.04 mag. Besides, from the top-right panel in the upper panel of Figure~\ref{fig:K2_A}, there is another significant peak beyond the 0.001 false alarm level (FAL) at the right side of the maximum 0.15511\,d peak, which makes the small peaks on either side of the main peak appear asymmetrically distributed.

\begin{figure}
	\includegraphics[width=\columnwidth]{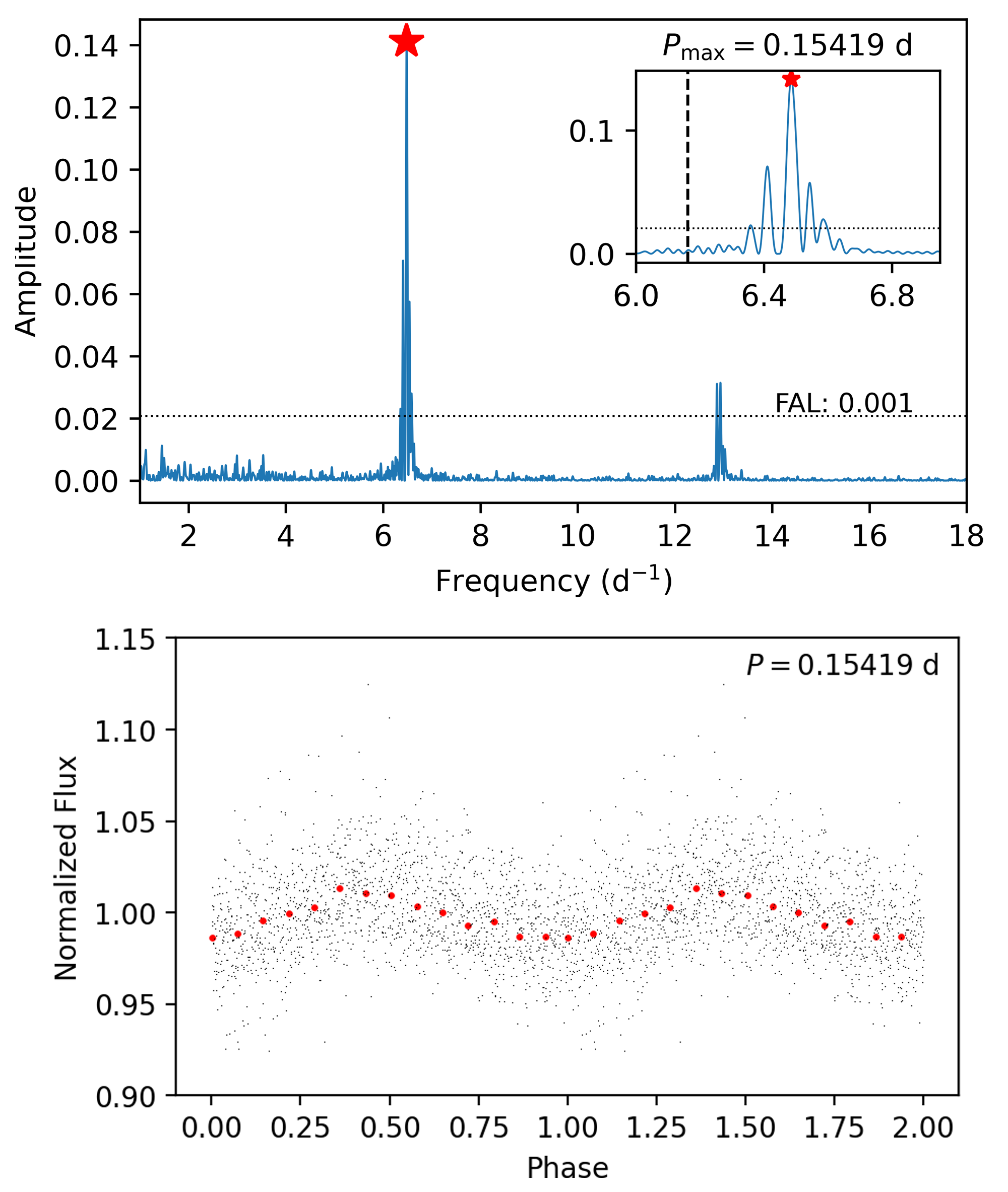}
	\caption{Upper panel: The Lomb-Scargle periodogram after the pre-whiten procedure to remove the 0.15511\,d signal and its second harmonic from the flattened K2 SFF light curve is calculated. The red star marks the maximum peak which has a period of 0.15419\,d. The black dotted line is 0.001 FAL. The right upper panel is a zoomed view near the maximum peak, and the vertical dashed line is the 0.16227\,d period obtained from the TESS data.
	Lower panel: Phase-folded light curve of the flattened and pre-whitened Kepler data is plotted using the period of 0.15419\,d. The gray dots are original data and the red dots are 16-minute binned data. For clarity, two phases are drawn repeatedly.}
	\label{fig:K2_B}
\end{figure}

To resolve the second peak at the right side of the maximum peak, we use the pre-whiten procedure. That is, we subtract the Lomb-Scargle model using the maximum 0.15511\,d period and its second harmonic from the original light curve, and then calculate a periodogram again for the residual. The result is shown in the upper panel of Figure~\ref{fig:K2_B}. The second period in the light curve is 0.15419\,d, and the side peaks caused by the window-function effect around this period are symmetrically distributed. Using this period to fit the residual light curve, we obtain the phase profile shown in the lower panel of Figure~\ref{fig:K2_B}. The amplitude of this profile is about 0.01 mag. The shape of the binned light curve also deviates from a sinusoidal form.

\begin{figure}
	\includegraphics[width=\columnwidth]{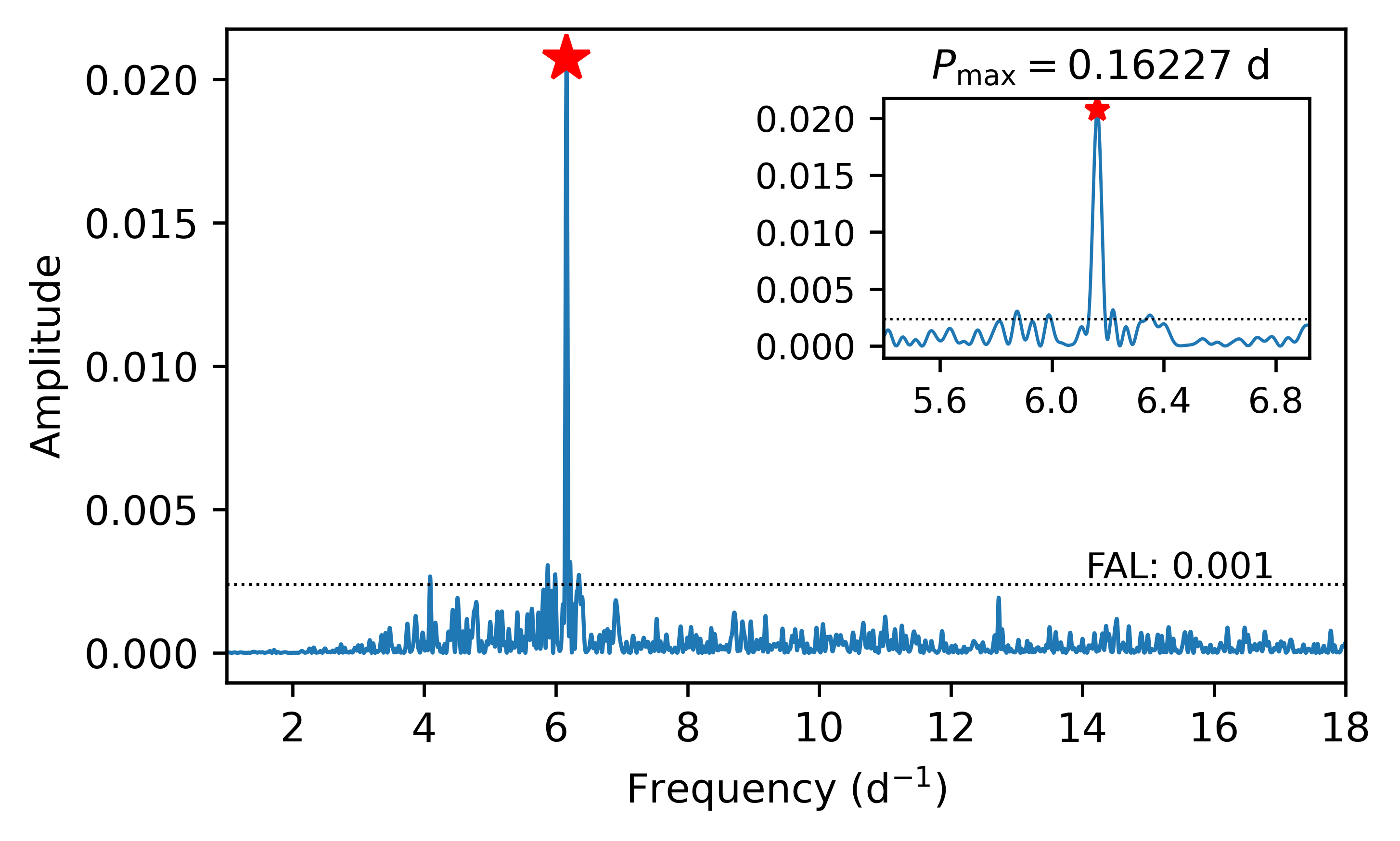}
	\caption{The Lomb-Scargle periodogram of the flattened TESS light curve is calculated. The red star marks the maximum peak which has a period of 0.16227\,d. The black dotted line is 0.001 FAL. The right upper panel is a zoomed view near the maximum peak.}
	\label{fig:TESS_p1}
\end{figure}

\begin{figure}
	\includegraphics[width=\columnwidth]{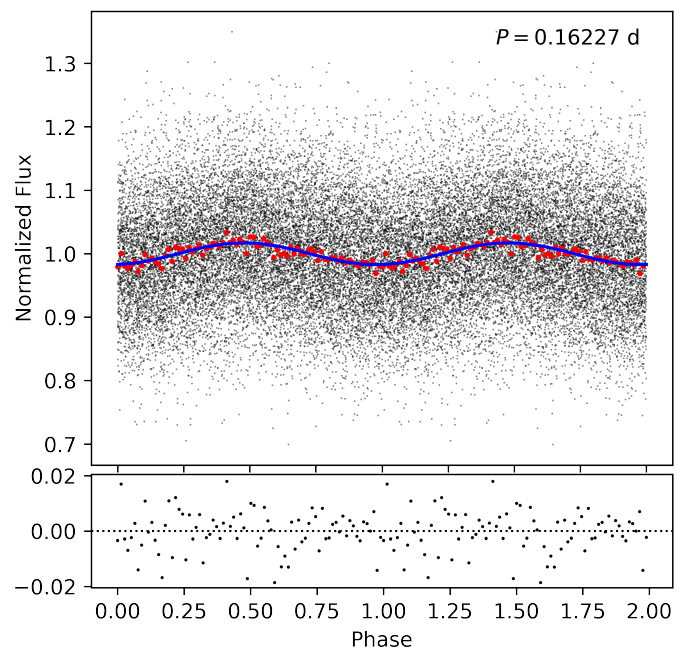}
	\caption{Phase-folded light curve of the flattened TESS data is plotted using the period of 0.16227\,d. The gray dots are original data and the red dots are 3-minute binned data. The blue curve is a sinusoidal function fitted by the least square method of the binned data. The lower panel plots the residuals between the sinusoidal function and the binned data. For clarity, two phases are drawn repeatedly.}
	\label{fig:TESS_bin}
\end{figure} % THE FIGURE NEED TO BE FIXED

\begin{figure}
	\includegraphics[width=\columnwidth]{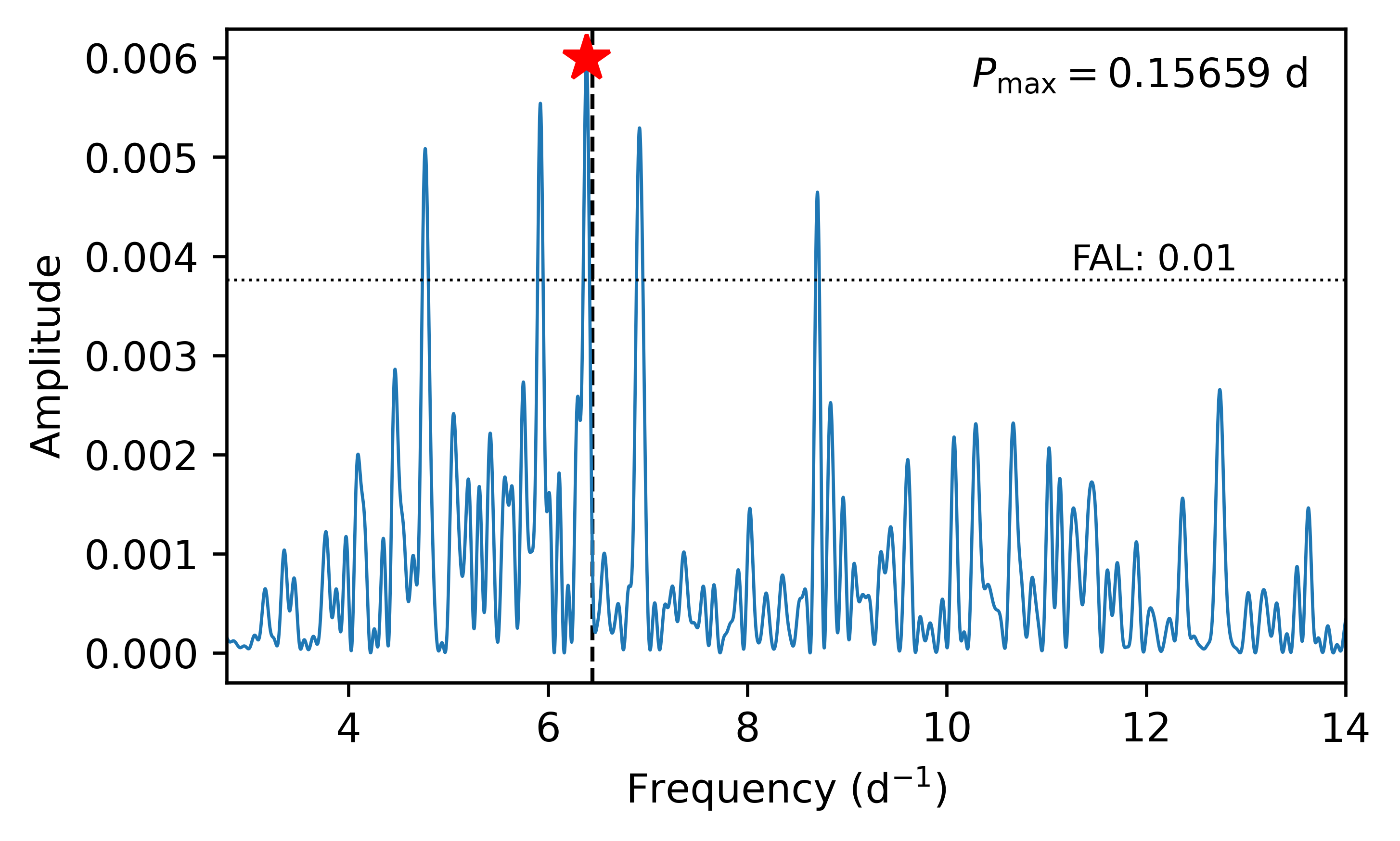}
	\caption{The Lomb-Scargle periodogram of the flattened and pre-whitened TESS light curve is calculated. The red star marks the maximum peak which has a period of 0.15659\,d. The black dotted line is 0.01 FAL. The vertical dashed line marks the 0.15511\,d period.}
	\label{fig:TESS_p2}
\end{figure}	
\subsubsection{TESS light curve}\label{sec:TESS_lc}
As seen in Panel 5 of Figure~\ref{fig:lc1}, the TESS observation is just at the standstill state prior to the large outburst 7. The light curve consists of both short time-scale changes and large-scale fluctuations. The large-scale fluctuations have a period of several days and are also revealed by the ASAS-SN g-band data after we shift and scale the relative flux of the TESS data for comparison. To reveal the short-term change, we flatten the light curve to remove the fluctuations on the scale of days and perform the Lomb-Scargle periodogram calculation. The result is shown in Figure~\ref{fig:TESS_p1}, from which a prominent peak of the period of 0.16227\,d is seen. After folding the flattened light curve on this period, the phase-folded light curve shown in Figure~\ref{fig:TESS_bin} reveals a sinusoidal shape. For comparison, the residuals between the sinusoids fitted by the least square method and the binned data are plotted in the lower panel. The deviation from a sinusoidal function is small.

By dividing the TESS light curve into segments and calculating the Lomb-Scagle periodogram for each of them, we discover a weak peak near the maximum peak in the first half of the TESS light curve. We perform the pre-whiten procedure to remove the 0.16227\,d sinusoidal signal and calculate a periodogram for the residual, a peak with a period of 0.15659\,d is revealed as seen in Figure~\ref{fig:TESS_p2}. The binned light curve folded by 0.15659\,d is not sinusoidal. This period is the most prominent peak in the periodogram with siginificance above 0.01 FAL, although four neighboring peaks also higher than this threshold without clear origins. The vertical black dashed line in Figure~\ref{fig:TESS_p2} denotes the period of 0.15511\,d derived from the Kepler data, which is very close to the 0.15659\,d peak. 

Finally, we calculate the periodogram for the whole TESS data and find that the large-scale fluctuations have a period of around 5.5\,d, or frequency of around 0.18 $\rm d^{-1}$ (see Figure~\ref{fig:TESS-long}). The Lomb-Scargle periodogram shows a wide hump. The large-scale fluctuations are not so stable in period as compared to a sinusoidal curve.

\subsection{Orbital period and negative superhump}\label{sec:periods}

We summarize all the periods that have been analyzed above from K2SFF and TESS data in Table~\ref{tab:periods}. We adopt the bootstrapping method described by \citet{2014MNRAS.437..510C} to determine the statistical errors on these periods.
The amplitudes are half of the difference between the maximum and minimum of the phase-folded profiles. $P_{k1}$, $P_{k2}$, and $P_{t2}$ are close and the phase-folded profiles of the three periods are not sinusoidal. These three periods seem to be not quite stable. The signal of $P_{t1}$ is sinusoidal. We discover that if we compare $P_{k1}$, $P_{k2}$, or $P_{t2}$ with $P_{t1}$, the differences are around a few percent, which reminds us of the characteristic of the existence of superhumps. The \textasciitilde 5\,d quasi-periodic fluctuations that arose in TESS light curve are also consistent with the frequency difference between $P_{t1}$ and $P_{t2}$. We suspect that the difference between $P_{k1}$ and $P_{k2}$ could be affected by the inaccurate correction of the satellite pointing issue from the original K2 light curve. 

\begin{figure}
	\includegraphics[width=\columnwidth]{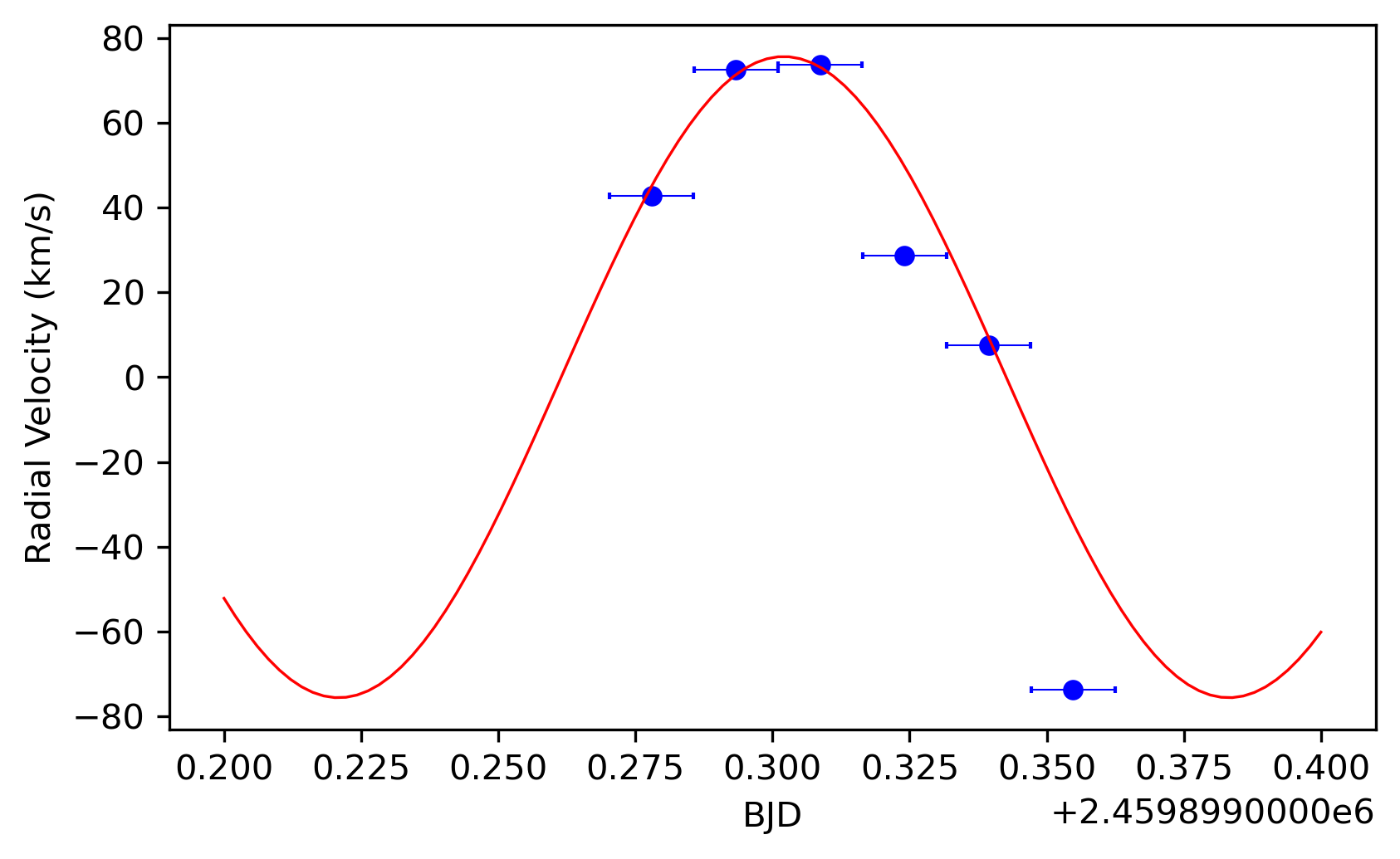}
	\caption{The radial velocity data of H$\alpha$ emission line obtained by Xinglong 2.16m telescope are plotted by blue points. Each half of the horizontal error bars indicates half the exposure time of each spectrum. The red curve is a sinusoidal curve with a fixed period of 0.16227\,d and an amplitude of 76 km/s.}
	\label{fig:rv} % NEED TO BE CHECKED AGAIN
\end{figure}

We use Gaussian fitting to obtain the central wavelengths of the H$\alpha$ emission line and convert them into radial velocities. The resulting H$\alpha$ radial velocities are shown in Figure~\ref{fig:rv} (a median velocity has been subtracted). Using the 0.16227\,d period derived from TESS data, we plotted a sinusoidal curve (red) to fit the radial velocity points. We find that the data points 1, 2, 3, and 5 fit the fixed period sinusoidal curve well, so we only choose to fit these four points. Because there are not enough radial velocity data, we cannot fully determine the exact radial velocity period and the phase, thus we can not judge which period in K2 and TESS data is orbital. However, from the current data, we rule out the possibility that twice the period from K2 and TESS light curves is the orbital period.

We therefore assume that $P_{t1}$, which is 3.89 h, corresponds to the orbital period, while $P_{k1}$, $P_{k2}$, and $P_{t2}$ reflect the periods of negative superhumps which are shorter than the orbital period. We have split the K2 light curve into small segments and calculated the periodogram for each one. We found that the peak have a medium fluctuations between 0.154 and 0.156 d and the double-peak feature is only present in the total periodogram. Therefore, $P_{k1}$ and $P_{k2}$ are not likely to be present at the same time. We consider that the superhump may be not quite stable, or the SFF correction is not perfect as we found large period changes in the original K2 light curve.
$P_{t3}$ corresponds to the nodal precession period of the tilted accretion disk, which is the beat period associated with both the orbital period and negative superhump period \citep{wood_sph_2009}, satisfying:
\begin{equation}
	\frac{1}{P_{\rm{prec}}}=\frac{1}{P_{\rm{nsh}}}-\frac{1}{P_{\rm{orb}}},
	\label{eq:beat}
\end{equation}
 where $P_{\rm{prec}}$ is the disk precession period, $P_{\rm{nsh}}$ is the negative superhump period, and $P_{\rm{orb}}$ is the orbital period. In the following section, we will justify the rationality of this assumption.

\begin{table*}
	\centering
	\caption{All periods derived from K2SFF and TESS data are summarized.}
	\label{tab:periods}
	\begin{tabular}{lcccr} % four columns, alignment for each
		\hline
		Data & Period & Frequency & Amplitude & Feature\\
		\hline
		
		K2SFF & $P_{k1}$ = 0.15511(1) d & $f_{k1}$ = 6.4470(6) d$^{-1}$ & 0.034 & Non-sinusoidal \\
		K2SFF & $P_{k2}$ = 0.15419(5) d & $f_{k2}$ = 6.4853(21) d$^{-1}$ & 0.013 & Non-sinusoidal \\
		TESS & $P_{t1}$ = 0.16227(4) d & $f_{t1}$ = 6.1626(13) d$^{-1}$ & 0.020 & Sinusoidal \\
		TESS & $P_{t2}$ = 0.15659(32) d & $f_{t2}$ = 6.386(13) d$^{-1}$ & 0.011 & Non-sinusoidal \\
		TESS & $P_{t3}$ = 5.47(2) d & $f_{t3}$ = 0.1829(5) d$^{-1}$& 0.063 & 
		\\
		\hline
	\end{tabular}
\end{table*}

\subsection{Mass ratio}\label{sec:q}
If the negative superhump period and the orbital period are assumed, the mass ratio of the system can be estimated. We utilize the simulation results by \citet{wood_sph_2009} and \citet{thomas_emergence_2015}, where they assume a white dwarf mass of 0.8 M$\odot$ and a mass-radius relation for the secondary. The negative excess $\epsilon_-$ is defined as:
\begin{equation}
	\epsilon_-=\frac{P_{\rm{nsh}}-P_{\rm{orb}}}{P_{\rm{orb}}},
	\label{eq:1}
\end{equation}
and the relation of $\epsilon_-$ and mass ratio $q$ is:
\begin{equation}
	\epsilon_-=-0.02263q^{1/2}-0.277q+0.471q^{3/2}-0.249q^2.
	\label{eq:2}
\end{equation}

As the signal of $P_{k1}$ is much stronger than that of $P_{k2}$ or $P_{t2}$, we take $P_{k1}$ as the negative superhump period and $P_{t1}$ as the orbital period. Then, the negative excess is calculated to be $\epsilon_-=-0.0441$, and the inferred mass ratio from Equation~\ref{eq:2} is 0.366. On the contrary, if we assume the positive superhump case, which is assuming the longer $P_{t1}$ as the period of positive superhumps and the shorter $P_{k1}$ as the orbital period, according to Equation~12 in \citet{wood_sph_2009}, the positive period excess $\epsilon_+$ is calculated to be 0.0462. This value requires $q=0.16$. However, even if the white dwarf were as large as 1.4 $M_{\odot}$, the secondary should be less than 0.2 $M_{\odot}$, which would be difficult to accommodate with the assumed 3.7 h orbital period. Therefore, only the negative superhump case is allowed and $P_{k1}$ can not be the orbital period.

We also estimate the uncertainty on $\epsilon_-$ and $q$. From the errors of $P_{k1}$ and $P_{t1}$ presented in Table~\ref{tab:periods}, we use the Monte Carlo method assuming a Gaussian error to estimate the uncertainty on $\epsilon_-$ to be 0.0002, and the uncertainty on $q$ to be 0.004.

\citet{zorotovic2011post} has revealed the white dwarf mass distribution among CVs, which shows that the average white dwarf mass in CVs is 0.83 $M_{\odot}$ and the highest proportion is around 0.8 $M_{\odot}$. If we assume a white dwarf mass as 0.83 $M_{\odot}$, the secondary mass should be 0.30 $M_{\odot}$ (effective temperature 3436 K, spectral type M3.5, and absolute V magnitude 10.84). According to the CV evolution model proposed by \citet{knigge2011evolution}, this secondary mass corresponds to an orbital period of 3.9 h, or 0.1625\,d, which is in coincidence with the assumed orbital period $P_{t1}$. This calculation makes it more reasonable that $P_{t1}$ is the true orbital period. In addition, it is impossible that $P_{t1}$ is the orbital period because the mass ratio derived from the positive excess in this condition would be too small and requires an impossibly large white dwarf mass.  

Moreover, if we otherwise assume a negative superhump period as $P_{k2}$ or $P_{t2}$, the corresponding negative period excess should be -0.0498 and -0.0350, respectively. The mass ratios are then derived as 0.49 and 0.22, respectively. For an orbital period of 3.9 h and the secondary mass of 0.30 $M_{\odot}$, the inferred white dwarf mass should be 0.61 and 1.36 $M_{\odot}$, respectively, which are both reasonable values under the Chandrasekhar limit. 

We propose two possible causes to explain the absence of the orbital signal in the K2 light curve. Firstly, the passband of Kepler photometry covers a large range in the optical band and is more sensitive at 600 nm than the near-infrared region \citep{borucki2016kepler}, while the passband of TESS covers from 600 nm to longer than 1000 nm, in which the radiation of the secondary is more effective and the contribution of the disk and the white dwarf is reduced. Secondarily, J0652+2436 is in an outburst phase as we can see from Figure~\ref{fig:lck2}, while during TESS observation the star did not show outbursts. The outbursting disk may outshine other components of the system and make the orbital signal disappear. The differences in the negative superhump period found in TESS and K2 data may originate from the instability of the disk precession. The systematics that underlies the K2 data may also contribute to an unstable period. Besides, $P_{t3}$ could be affected by the instability of the response of the TESS instrument.

\subsection{Tidal resonance}\label{sec:resonance}
Using the derived mass ratio, the tidal limitation of the accretion disk can be determined. This limitation stands for the maximum radius that the disk can maintain circular. According to \citet{warner_cataclysmic_1995}, the tidal limitation can be calculated with the following equation for $0.03\textless q\textless1$:
\begin{equation}
	r_{\rm{tidal}}=\frac{0.60}{1+q},
	\label{eq:tidal_radius}
\end{equation}
where $r_{\rm{tidal}}$ provide radii in unis of the orbital separation. Plugging $q = 0.366$ to the equation, we determine the $r_{\rm{tidal}}$ to be 0.44. Thus, if the outer part of the disk extends beyond this limitation, the trajectories of the disk particles will intersect each other and generate dissipation.

In the smoothed particle hydrodynamics (SPH) simulations of the accretion disk by \citet{wood_sph_2009}, they found positive superhumps developed for mass ratio up to 0.35, so that $q=0.366$ is marginally possible for the tidal instability to set in. As we see in Figure~\ref{fig:outbursts} that the large outbursts 1-7 (except outburst 8) were triggered at almost identical brightness, we propose that the observed large outbursts may be caused when the radius of the growing accretion disk increases to reach the 3:1 resonance radius. Before these large outbursts shown in Figure~\ref{fig:outbursts}, the average brightness was gradually increasing. Also, during the standstill before outburst 7, as described in Section~\ref{sec:outbursts} , there is a steady increase in brightness. These slowly rising processes give support the increasing radius of the accretion disk. 

\section{Discussion}\label{sec:discussion}

\subsection{IW And stars on Gaia CMD}\label{sec:cmd}

\begin{figure}
	\includegraphics[width=0.5\textwidth]{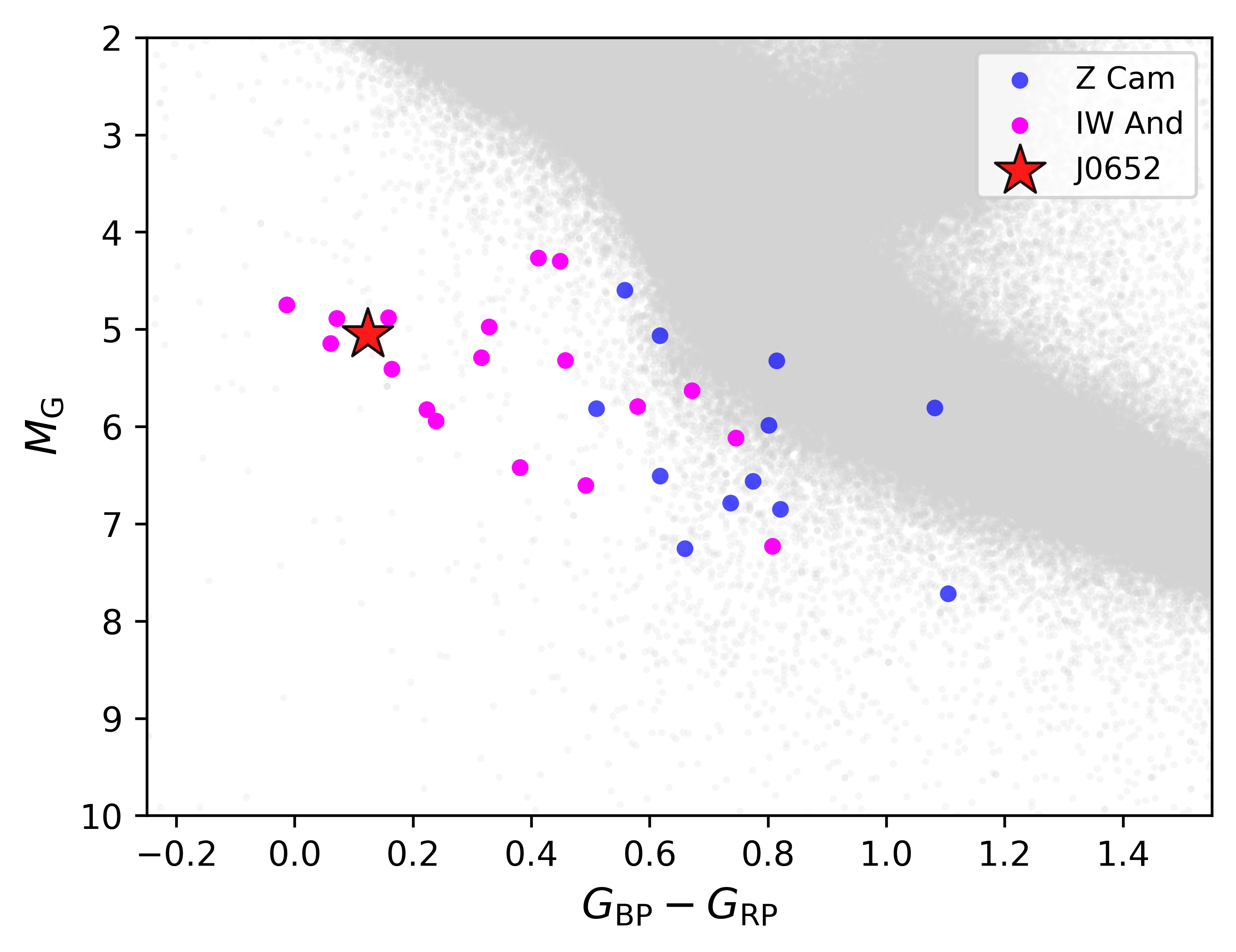}
	\caption{The distribution for Z Cam-type (blue) and 18 known IW And-type (magenta) dwarf novae are plotted on the Gaia CMD. J0652+2436 is marked by the red star. 
	Stellar objects (grey) randomly selected from LAMOST DR9 spectral database are also plotted for comparison. }
	\label{fig:cmd}
\end{figure}

In Figure~\ref{fig:cmd}, 18 IW And-type stars that have been reported are plotted on the Gaia colour-magnitude diagram (CMD), which include
IW And, V513 Cas, HX Peg, AH Her, AT Cnc,  
V507 Cyg, IM Eri, FY Vul, 
HO Pup, 
ST Cha, 
ASAS J071404+7004.3, 
KIC 9406652, 
BC Cas, 
BO Ceti, 
HL Cma, UZ Ser, MV Leo, and
ES Dra (candidate IW And-type star)
\citep{hameury2014anomalous,kato2019three, kato2022long,lee2021ho,kato2021nature,kato2022analysis,kimura2020kic,kato2020bc,kato2021bo,ZCamstarsinthe21,kato2022negative}. J0652+2436 is plotted as the red star in the figure. The absolute Gaia magnitudes $M_{\rm{G}}$ and the Gaia BP - RP colour are calculated from Gaia early data release 3 (EDR3) \citep{prusti2016gaia, brown2021gaia}, using the geometrical distances from \citet{bailer2021estimating} and the code of three-dimension dust map by \citet{guo2021three} to calculate E(B-V). The magnitudes are the mean values of several Gaia measurements. The exact colour and Gaia absolute magnitudes are listed in Table~\ref{tab:Gaia CMD}.

Among these 18 known IW And-type stars, we confirme that 17 stars show either standstills terminated by outbursts, continuous small outbursts happening during standstill states, or both after we checked their ASAS-SN and ZTF data. Except for UZ Ser, which was reported by \citet{ZCamstarsinthe21} from the AAVSO International Database. It did not show noticeable standstills in the data we checked.

Putting all 18 known IW And stars on the CMD, it is interesting to find that they are located in the bluer region among Z Cam stars. The bluer Z Cam stars on Gaia CMD are almost all IW And stars. This probably reflects a common property among IW And stars. In particular, J0652+2436 is among the bluest IW And stars. However, the sparse sampling of Gaia photometry and the significant light variations of these stars make it difficult to reveal the true average brightness between IW And and Z Cam stars. Further study of the properties of IW And stars and more samples to be discovered in the future large-scale photometry surveys will provide more evidence for the difference between IW And stars and Z Cam stars. Comparing the number of known IW And stars with that of Z Cam stars, we also notice the high incidence of IW And phenomena among Z Cam stars.

%IW And, V513 Cas, HX Peg, AH Her, AT Cnc,  \citep{hameury2014anomalous}
%V507 Cyg, IM Eri, FY Vul, \citep{kato2019three, kato2022long}
%HO Pup, \citep{lee2021ho}
%ST Cha, \citep{kato2021nature}
%ASAS J071404+7004.3, \citep{kato2022analysis}
%KIC 9406652, \citep{kimura2020kic}
%BC Cas, \citep{kato2020bc}
%BO Ceti, \citep{kato2021bo}
%HL Cma, UZ Ser, MV Leo, \citep{ZCamstarsinthe21}
%ES Dra (candidate IW And star) \citep{kato2022negative}.

%\textcolor{red}{\subsection{Absolute magnitudes in different states (deleted)}}

\subsection{Comparison with NY Ser and BO Cet}\label{sec:compare}

J0652+2436 might be the first IW And-type dwarf nova exhibiting both negative superhump and tidal instability. \citet{kato2022analysis} pointed out that a tilted disk is probably not responsible for IW And phenomenon, as negative superhumps have not been reported in most IW and stars and the tilted-disk model can not explain the case of KIC 9406652. We suggest that the outbursts in J0652+2436 may have been caused by an increase in the disk radius, and a tilted disk could potentially contribute to the more complex outburst morphology observed in this star, which distinguishes it from other known IW And stars.

It is noticeable that the largest several outbursts in J0652+2436, after which the star entered lower states and generated a series of shorter outbursts, are quite similar to that of SU UMa-type dwarf nova NY Ser. NY Ser shows superoutbursts started from standstills as reported by \citet{kato2019three}. They proposed that the superoutbursts observed in NY Ser provide evidence of disk expansion, which ultimately leads to tidal instability. They further suggest that IW And-type stars may experience a similar situation. During the standstill state, the accretion disk expands, while the inner region of the disk remains hot and stable until tidal instability is initiated at the 3:1 resonance radius. Subsequently, the instability spreads from the outer region to the inner region of the disk, triggering an outburst.
This is likely the scenario present in J0652+2436. Prior to the outbursts, J0652+2436 typically undergoes a gradual brightening process, which could indicate the expansion of the accretion disk. As shown in Figure~\ref{fig:outbursts}, after the gradual brightening process, the outbursts started from nearly identical magnitude, which proves that the same condition is reached. This might suggest that the accretion disk expands to the limit at the tidal instability starting to trigger. 

J0652+2436 exhibits more complex outburst characteristics compared to other stars showing IW And-type phenomena. Specifically, during the intermediate bright state (magnitude 15.5-16), there are not only outbursts that terminate slowly rising standstill states and "heart-beat"-like outbursts, but also long periods of continuous low-amplitude outbursts without interval. Additionally, during the lower states triggered by large outbursts, there are smaller outbursts with shorter recurrence times and durations. We propose that the differences in recurrence time, amplitude, profile, and termination of outbursts from the intermediate states are due to the different conditions when the accretion disk reaches the tidal limitation. Considering the precession of the tilted disk in J0652+2436, it is likely that additional complexity is introduced to these conditions. We speculate that after large outbursts 2 and 7, when J0652+2436 entered deep low states, the disk may have been significantly depleted. In contrast, after large outbursts 3, 5, and 6, J0652+2436 did not enter the low state, indicating that the disk was not fully emptied. Further investigation into the dynamics of the accretion disk in J0652+2436 is necessary, and it still remains unclear whether the disk expansion occurs in all IW And-type stars.

\citet{kato2021bo} reported that BO Cet is another dwarf nova showing both IW And-type outbursts and SU UMa-type positive superhump features. They have analyzed that BO Cet has a mass ratio close to J0652+2436, 0.31-0.34, which makes the disk of BO Cet able to reach the 3:1 resonance radius. But BO Cet is a typical SU UMa-type dwarf nova that shows positive superhumps. However, we did not identify positive superhumps in the current data of J0652+2436.

\section{Conclusions}\label{sec:conclusions}

J0652+2436 displays diverse light variations including characteristics of standstills followed by outbursts and ‘heart-beat’-like outburst-dip sequences, leading us to classify this object as a new IW And-type dwarf nova. Fortunately, the TESS observation coincided with the a prolonged standstill phase, allowing us to observe significant brightness fluctuations with an approximate period of \textasciitilde5\,d, which is indicative of disk precession. An analysis of the TESS light curve reveals a distinct sinusoidal signal with a period of 0.16227\,d, while both the K2 and TESS light curves contain signals near 0.155\,d. We posit that the 0.16227\,d period likely corresponds to the orbital period, whereas the 0.155\,d signal is attributed to negative superhumps, suggesting the presence of a tilted and precessing disk. Through the examination of emission-line radial velocity data, we exclude the possibility of the orbital period being twice the value of 0.16227\,d, although further radial velocity observations are necessary to definitively confirm the true orbital period.
 
 The detection of negative superhumps enables the measurement of the mass ratio. After calculating the superhump negative excess, we determined that the mass ratio is 0.366, which make the tidal dissipation possible to happen when the disk radius grows. The gradual rising brightness before outbursts could be evidence of an expanding accretion disk, in favor of the conjecture proposed by \citet{kato2019three}. This may explain the frequent outbursts in the intermediate bright state and the outbursts at the end of the standstill states. Therefore, J0652+2436 is possibly the first IW And-type dwarf nova having a tilted disk generating tidal instability. A more thorough investigation of the dynamics of the accretion disk in J0652+2436 is necessary to fully comprehend the relationship between disk expansion, IW And phenomena, and the impact of the tilted disk. Therefore, J0652+2436 serves as a crucial research laboratory for exploring the dynamics of the accretion disk and enhancing our understanding of IW And phenomena.

 Upon analyzing the Gaia CMD, we have identified a distinctive characteristic pertaining to the distribution of the known 18 IW And stars. It is observed that these stars are typically located in the bluer region compared to normal Z Cam stars. Notably, J0652+2436 can be clearly classified as a member of the IW And star population on the Gaia CMD. The pattern of distribution suggests the presence of a unique common property inherent to IW And stars themselves. Future surveys that identify a greater number of IW And stars will enable us to substantively compare the difference of the luminosities of IW And versus Z Cam stars and find out whether IW And stars have higher mass-transfer rate as shown in \citet{Inight2023}. 
It is important to highlight that high-cadence and long-time monitoring of these systems is essential to capture the lowest and brightest states, and to uncover the rapid changes in accretion disks, allowing for a better comprehension of their physical properties.

% The last numbered section should briefly summarise what has been done, and describe the final conclusions which the authors draw from their work.

\section*{Acknowledgements}

 This work is supported by the National Natural Science Foundation of China (NSFC) under Grant 12173013, 12090044. This work was also supported by the Natural Science Foundation of Hebei Province under Grant A2021205006 and by the project of Hebei provincial department of science and technology under Grant Number 226Z7604G. W.C. is also supported by the China Manned Space Project. This work is partly supported by the Innovation Project of Beijing Academy of Science and Technology (23CB061). 
 
 We acknowledge the support of the staff of the Xinglong 2.16 m telescope. LAMOST is a National Major Scientific Project built by the Chinese Academy of Sciences, which has been provided by the National Development and Reform Commission.
 %LAMOST is operated and managed by National Astronomical Observatories, Chinese Academy of Sciences.
  
 This research made use of \textsc{lightkurve}, a Python package for Kepler and TESS data analysis \citep{2018ascl.soft12013L}. This work made use of \textsc{astropy}:\footnote{http://www.astropy.org} a community-developed core Python package and an ecosystem of tools and resources for astronomy \citep{astropy:2013, astropy:2018, astropy:2022}.  This work has made use of data from the European Space Agency (ESA) mission
 {\it Gaia} (\url{https://www.cosmos.esa.int/gaia}), processed by the {\it Gaia} Data Processing and Analysis Consortium (DPAC, \url{https://www.cosmos.esa.int/web/gaia/dpac/consortium}). Funding for the DPAC has been provided by national institutions, in particular the institutions participating in the {\it Gaia} Multilateral Agreement.
%%%%%%%%%%%%%%%%%%%%%%%%%%%%%%%%%%%%%%%%%%%%%%%%%%
%\section*{Data Availability}

% The inclusion of a Data Availability Statement is a requirement for articles published in MNRAS. Data Availability Statements provide a standardised format for readers to understand the availability of data underlying the research results described in the article. The statement may refer to original data generated in the course of the study or to third-party data analysed in the article. The statement should describe and provide means of access, where possible, by linking to the data or providing the required accession numbers for the relevant databases or DOIs.

\section*{Data Availability}
The data underlying this article will be shared on reasonable request to the corresponding author.



\bibliographystyle{mnras}
\bibliography{bibliography}

%\begin{thebibliography}{99}
%\end{thebibliography}

%%%%%%%%%%%%%%%%%%%%%%%%%%%%%%%%%%%%%%%%%%%%%%%%%%

%%%%%%%%%%%%%%%%% APPENDICES %%%%%%%%%%%%%%%%%%%%%

\appendix

\section{Some extra material}

\begin{figure*}
	\includegraphics[width=\textwidth]{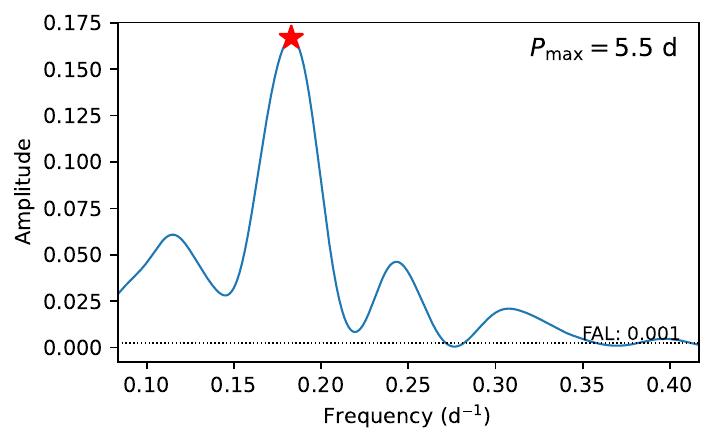}
	\includegraphics[width=\textwidth]{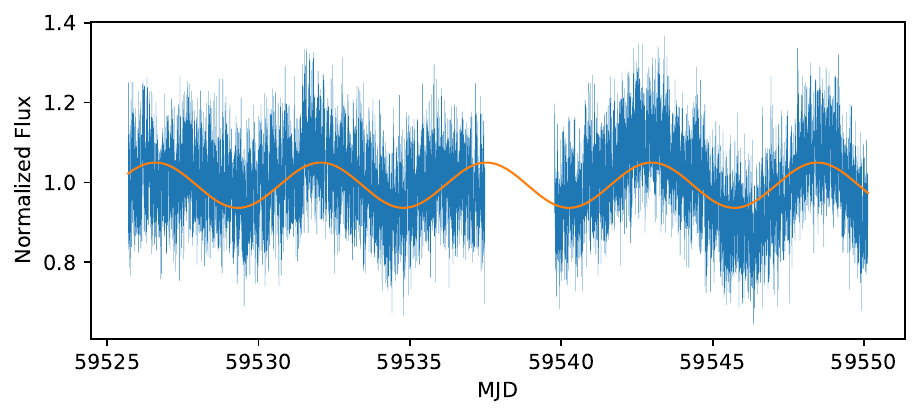}
	\caption{Upper panel: The Lomb-Scargle periodogram of the TESS Sector 45 light curve is calculated. The red star marks the maximum peak which has a period of 5.5\,d. Lower panel: The Lomb-Scargle model with 5.5\,d period is plotted in orange and the TESS data is plotted in blue.} 
	\label{fig:TESS-long}
\end{figure*}

\begin{table*}
	\caption{The Gaia CMD parameters for the studied stars. The distances are taken from \citet{bailer2021estimating} and the E(B-V) values are calculated through the code from \citet{guo2021three}. Except for ASAS J071404+7004.3, the orbital periods are taken from the 2017 edition catalog by \citet{ritter2003catalogue}.}
	\centering
	\scriptsize
	\label{tab:Gaia CMD}
	\begin{tabular}{llllllllllllllllll}
		\hline
		Name & Type$^{\rm a}$ & Distance(pc) & E(B-V) & BP-RP & MG & Orbital Period(d)\\ 
		\hline
        IW And & IW & 849.6 & 0.060 & -0.01 & 4.75 &0.155&\\ 
		V507 Cyg & IW & 604.0 & 0.179 & 0.49 & 6.61 & unknown&\\ 
		IM Eri & IW & 185.5 & 0 & 0.16 & 5.41 &0.145635&\\ 
		FY Vul & IW & 573.0 & 0.119 & 0.32 & 5.29 & unknown&\\ 
		HO Pup & IW & 624.5 & 0.038 & 0.16 & 4.88 & unknown&\\ 
		ASAS J071404+7004.3 & IW & 210.2 & 0.022 & 0.06 & 5.15 &0.1366454$^{\rm b}$&\\ 
		ST Cha & IW & 689.9 & 0.103 & 0.07 & 4.89 &0.285&\\ 
		KIC 9406652 & IW & 342.1 & 0.011 & 0.33 & 4.98 &0.2544&\\ 
		BC Cas & IW & 1830.9 & 0.645 & 0.45 & 4.30 & unknown &\\ 
		V513 Cas & IW & 835.4 & 0.759 & 0.41 & 4.27 &0.217
		&\\ 
		HX Peg & IW & 585.7 & 0.048 & 0.46 & 5.32 &0.2008
		&\\ 
		AH Her & IW & 327.9 & 0.018 & 0.67 & 5.63 &0.258116
		&\\ 
		AT Cnc & IW & 455.0 & 0.033 & 0.75 & 6.12 &0.2011
		&\\ 
		BO Ceti & IW & 494.0 & 0.028 & 0.22 & 5.83 &0.1398
		&\\ 
		HL Cma & IW & 293.5 & 0.029 & 0.58 & 5.80 &0.216787
		&\\ 
		UZ Ser & IW & 308.8 & 0.037 & 0.81 & 7.23 &0.17589
		&\\ 
		MV Leo & IW & 999.1 & 0.021 & 0.38 & 6.42 &0.1461
		&\\ 
		ES Dra & IW & 680.2 & 0.022 & 0.24 & 5.94 &0.1766
		&\\ 
		LAMOST J0652+2436 & IW & 1183.9 & 0.039 & 0.12 & 5.06 &0.16227&\\ 
		NY Ser & SU & 808.6 & 0.045 & 0.24 & 6.47 &0.0978
		&\\ 
		HS 2325+8205 & ZC & 463.8 & 0.068 & 1.11 & 7.72 &0.194335
		&\\ 
		CRTS J220031.2+033430 & ZC & 3413.5 & 0.058 & 1.08 & 5.81 & 0.165&\\
		AY Psc & ZC & 715.3 & 0.051 & 0.82 & 6.85 &0.217321
		&\\ 
		BX Pup & ZC & 739.1 & 0.055 & 0.51 & 5.82 &0.127
		&\\ 
		EM Cyg & ZC & 355.9 & 0.020 & 0.82 & 5.33 &0.290909
		&\\ 
		PY Per & ZC & 523.7 & 0.033 & 0.66 & 7.26 &0.1548
		&\\ 
		RX And & ZC & 196.6 & 0 & 0.78 & 6.56 &0.209893
		&\\ 
		SY Cnc & ZC & 401.2 & 0.028 & 0.56 & 4.60 &0.382375
		&\\ 
		TZ Per & ZC & 456.9 & 0.085 & 0.62 & 5.07 &0.262906
		&\\ 
		VW Vul & ZC & 542.9 & 0.057 & 0.62 & 6.51 &0.1687
		&\\ 
		WW Cet & ZC & 218.1 & 0 & 0.74 & 6.79 &0.1758
		&\\ 
		Z Cam & ZC & 213.5 & 0.017 & 0.80 & 5.99 &0.289841
		&\\ 
		\hline
	\end{tabular}
\\
a. IW, SU and ZC stand for IW And-type, SU UMa-type and Z Cam-type dwarf novae, respectively. \\
b. The orbital period is taken from \citet{inight2022asas}. \\
\end{table*}

%		WW Cet & ZC & 15.10$^{\rm a}$ & 14.03 & 11.32 & 14.61 & 13.69 & 11.38 & %7.34 & 6.99 & 4.63 & 4.69 & 8.41 & 7.92 \\ 
%\end{tabular}
%\\ 
%a. The apparent magnitude may be inaccurate because there are not enough data %points. \\

% Don't change these lines
\bsp	% typesetting comment
\label{lastpage}
\end{document}